# Early warning signals for primary and secondary bifurcation to oscillatory instabilities

Rohit Radhakrishnan[1,2], Prasana Kumar[1,2], Induja Pavithran[1,2,a], and R. I. Sujith[1,2*]

[1] Department of Aerospace Engineering, Indian Institute of Technology Madras,
    Chennai, 600036, India
[2] Centre of Excellence for studying Critical Transitions in Complex Systems,
    Indian Institute of Technology Madras, Chennai 600036, India
[a] Present address: The Jacob Blaustein Institutes for Desert Research, Ben Gurion
                     University of the Negev

*Corresponding author: R. I. Sujith

Email: sujith@iitm.ac.in

**Keywords:** Early warning signals, Dynamical systems, Bifurcation theory, Natural visibility graph, primary and secondary bifurcations, oscillatory instabilities

**Abstract**

In several natural and engineering systems, changes in control parameters can trigger bifurcations that lead to sustained or growing periodic oscillations, indicating the onset of oscillatory instabilities. Such emergent behaviour often results from positive feedback between interacting subsystems, resulting in large-amplitude oscillations that can be detrimental. Several precursors are available to provide early warning of an impending oscillatory instability. In reality, practical systems may exhibit different sequences of bifurcations, including a primary bifurcation to an oscillatory state that may be either continuous or abrupt, followed by an abrupt secondary bifurcation, and further transitions beyond the secondary bifurcation. Existing precursors for oscillatory instabilities typically forewarn the onset of the primary bifurcation to an oscillatory state and tend to saturate once the system enters the oscillatory regime. Notably, primary bifurcations often involve lower amplitudes compared to the more severe states after secondary bifurcation. In this study, we propose a methodology based on spectral visibility graphs to get forewarning for both primary and secondary bifurcations. The method inherently captures the evolution of the harmonic content of the signal relative to other frequency components. The approach employs tuning of a single sensitivity parameter to detect different sequences of bifurcation. We demonstrate the usefulness of our method for thermo-acoustic and aero-acoustic instabilities in multiple engineering systems involving turbulent flow and reactions. Our methodology can help systems prepare in advance or even avoid undesirable transitions. Tuning the sensitivity parameter allows adaptive, risk-based warnings, ensuring high performance without tipping into undesirable regimes.

## Introduction

Oscillatory instabilities are widespread in various systems, often emerging from a state of low-amplitude aperiodic fluctuations (1). In engineering systems, these instabilities typically have a detrimental impact on both system performance and lifespan. In turbulent systems, self-organization driven by feedback between subsystems can give rise to oscillatory instabilities, as observed in thermoacoustic (2), aeroacoustic (3), and aeroelastic systems (4). These instabilities may manifest as low-amplitude oscillations, capable of causing fatigue, wear and tear of components and can even result in the failure of engine components (5) in the system, or as high-amplitude oscillations that can lead to catastrophic events in rocket engines (6, 7) and power plants (8).

Identifying precursors to oscillatory instabilities is crucial, as these instabilities can reach high amplitudes, leading to catastrophic events. Despite the unique characteristics of different systems, the dynamics near the bifurcation point exhibits universal behavior across various systems (9). Previous studies, have mainly focused on developing precursors for the primary bifurcation to oscillatory instabilities (i. e., predicting the onset of the transition to low-amplitude oscillations in the system) using various methods such as critical slowing down (10, 11), translational error (13, 14), symbolic time series analysis (15), fractal and multifractal characteristics of the time series (16), spectral measures (17), and deep learning algorithms (18–20).

In many practical turbulent systems, the system transitions from an aperiodic state of operation to an oscillatory state, referred to as thermoacoustic, aeroacoustic, or aeroelastic instability in the respective physical contexts. In general, the nonlinearities in the system influence the balance between energy input (driving) and energy dissipation (damping) in the system (21). The net energy balance leads to either a supercritical (smooth) or subcritical (abrupt) Hopf bifurcation of the acoustic subsystem to limit cycle oscillations (22). However, in certain cases, the influence of higher-order nonlinearities destabilizes a stable limit cycle, leading to a secondary fold bifurcation and the emergence of high-amplitude limit cycle oscillations (24). Such transitions have been reported in various systems, including aircraft flight dynamics, where they manifest as large-amplitude rolling oscillations known as wing rock (25), and in axial-flow compressors, where surge is an oscillatory instability that manifests as large-amplitude limit-cycle oscillations in mass flow rate and pressure rise (26). Ananthkrishnan et al. (27) theoretically predicted the occurrence of a secondary bifurcation leading to high-amplitude thermoacoustic instability, which was later confirmed experimentally in a laminar setup (28), in turbulent reacting flow systems such as an annular combustor (29), and in a turbulent axial combustor (30).

In the current study, we developed a methodology and a measure to forewarn the different sequences of bifurcation to oscillatory instabilities for practical systems. The measure can give a forewarning based on the different sequences of bifurcations to oscillatory instabilities. These may include a primary bifurcation resulting in low-amplitude oscillations or a primary bifurcation followed by an abrupt secondary transition to high-amplitude oscillations. We provide forewarning by constructing the natural visibility graph of the spectral domain of the experimentally obtained time series data of an appropriate variable. We use the acoustic pressure fluctuations as the variable for the thermoacoustic and aeroacoustic system studied in this work. The method relies on a single adjustable parameter, the '$q$ parameter', which enables forewarning for different bifurcation sequences. By providing early warnings prior to the onset of instabilities, this approach helps to maximise system performance and prevent the system from tipping into impending oscillatory instabilities.

**Traditional early warning measures for oscillatory instability**

We initially test the performance of traditional early warning measures such as variance, lag-1 autocorrelation, variance of autocorrelation, skewness, and kurtosis. These measures capture the critical slowing down before the system tips to another state, thereby forewarning an impending bifurcation to oscillatory instability. We consider the data from a turbulent reacting flow in an annular geometry where combustion ensues along with turbulent reactive flow in multiple tubes (29, 31), please refer to *Methods* Fig. 7. In our experiments, the turbulent reacting flow occurs in an annular passage between two cylindrical ducts, within which sixteen burner tubes are mounted. Each burner is equipped with an axial swirler, designed to impart a solid-body, counter-clockwise rotation to the flow downstream of the swirler (see *Methods* for details). The system uses premixed air and liquefied petroleum gas (LPG) as the fuel. We adjust the equivalence ratio ($\phi$), defined as the ratio between the actual fuel-air ratio and the stoichiometric fuel-air ratio. In this study, $\phi$ is varied in the range of 0.8 (fuel lean) to 1.1 (fuel rich) conditions. Acoustic pressure fluctuations measured using piezoelectric transducers were recorded for 3 s at a sampling rate of 10 kHz.

Figure 1 shows the variation of the root-mean-square (RMS) value of acoustic pressure fluctuations, as a function of the equivalence ratio ($\phi$), represented by blue-filled circles. For $\phi \lesssim 0.92$, the acoustic pressure fluctuations are aperiodic and exhibit amplitudes of the order of $10^1$ Pa, referred to as the state of safe operation or the state of combustion noise (29, 31). As $\phi$ increases, the system enters a regime characterized by low-amplitude periodic oscillations interspersed with very low-amplitude aperiodic behavior, known as intermittency (23). Beyond $\phi \approx 0.95$, the acoustic pressure gradually increases, with amplitudes reaching approximately $10^2$ Pa. We identify this region as low-amplitude thermoacoustic instability. At $\phi \approx 1$, a sharp increase in pressure amplitude is observed, indicating a transition from low to high-amplitude thermoacoustic instability, where pressure fluctuations reach $10^3$ Pa. Despite the sudden change in amplitude, the dominant frequency remains nearly constant at $f_s = 220 \pm 10$ Hz (shown in Fig. 4), which is close to the first longitudinal mode of the system (31).

The system exhibits a sequence of transitions to oscillatory instabilities, with a primary bifurcation from the state of safe operation → intermittency → low-amplitude thermoacoustic instability, followed by a secondary bifurcation from low-amplitude thermoacoustic instability → high-amplitude thermoacoustic instability. The transition to high-amplitude thermoacoustic instability results in an approximately 1000 Pa increase in the RMS value of sound intensity, corresponding to a 5-fold increase in amplitude compared to low-amplitude thermoacoustic instability. Both low and high-amplitude thermoacoustic instabilities are detrimental to engine performance and therefore undesirable. Early detection of these primary (safe operation → low-amplitude thermoacoustic instability) and secondary bifurcations (low to high-amplitude thermoacoustic instability) is critical for preventing the system from tipping into such undesirable states. Towards this, we initially consider the traditional measures as shown in Fig. 1 on the primary and secondary bifurcation data of the annular combustor. The variance (Fig. 1A), which indicates the deviation of the time series from its average, increases with the root-mean-square of acoustic pressure fluctuations.

The lag-1 autocorrelation (Fig. 1B) quantifies the correlation between the time series and a delayed version of itself by one sampling interval (0.1 ms in this study). In the present system, the lag-1 autocorrelation in the initial states (i.e., safe operation region) is approximately 0.8. This relatively high value can be attributed to the strong temporal correlations inherent to the system dynamics, combined with the high sampling frequency of 10 kHz employed in the measurements. At this sampling rate, the temporal separation between consecutive samples is small, resulting in significant correlation between adjacent data points. Lag-1 autocorrelation increases as we transition to the state of intermittency. As the system attains the state of low-amplitude thermoacoustic instability, the lag-1 autocorrelation tends to a value of unity and remains there even during the high-amplitude thermoacoustic instability. The value of lag-1 autocorrelation tending to a value of unity is expected, since the time series for both low and high-amplitude thermoacoustic instability are close to a state of periodic oscillation.

The variance of autocorrelation (Fig. 1C) is a measure that quantifies the variation of the autocorrelation across different lags. Here, we used up to 94 lags (equivalent to two acoustic cycles since each cycle contains 47 sample points) for computing the variance of autocorrelation (11). The variance of autocorrelation grows in value well before the onset of intermittency, giving an early warning for the primary bifurcation to the low-amplitude thermoacoustic instability. However, as the system reaches the low-amplitude thermoacoustic instability, the value becomes closer to 0.5, which is the value expected for a periodic oscillation (11). The value of the variance of autocorrelation saturates at the low-amplitude thermoacoustic instability and remains the same throughout, even during the high-amplitude thermoacoustic instability.

The skewness and kurtosis are the normalised third and fourth central moments of the distribution of the time series data. The value of skewness (Fig. 1D) exhibits a lot of scatter up to the low-amplitude thermoacoustic instability and then increases in value at the transition point to high-amplitude thermoacoustic instability. Kurtosis (Fig. 1E) remains close to a value of three during the safe operation region and gradually decreases and reaches a value of 1.5 during low-amplitude thermoacoustic instability, and then remains at that value even at the state of high-amplitude thermoacoustic instability.

The Hurst exponent is a measure that represents the temporal correlation present in the time series data collected from the system, as well as the emergence of periodicity. Hurst exponent is two - fractal dimension. Hurst exponent (Fig. 1F) is close to 0.2, and as the system approaches low-amplitude thermoacoustic instability, it approaches a value close to zero. Hurst exponent remains close to zero from low-amplitude thermoacoustic instability to high-amplitude thermoacoustic instability. We observe that lag-1 autocorrelation, variance of autocorrelation, kurtosis, and Hurst exponent (shown in Fig. 1) can provide an early warning for an impending primary bifurcation (i.e., safe operation → low-amplitude thermoacoustic instability). The value of skewness exhibits a sudden change in value at the point of transition to secondary bifurcation (low-amplitude thermoacoustic instability → high-amplitude thermoacoustic instability). However, these measures are able to forewarn only one of the transitions - either primary or secondary bifurcation. Thus, there is a need for new measures capable of providing early warnings for the primary and secondary bifurcation, and demarcating the transition to high-amplitude oscillations that can be detrimental to the system.

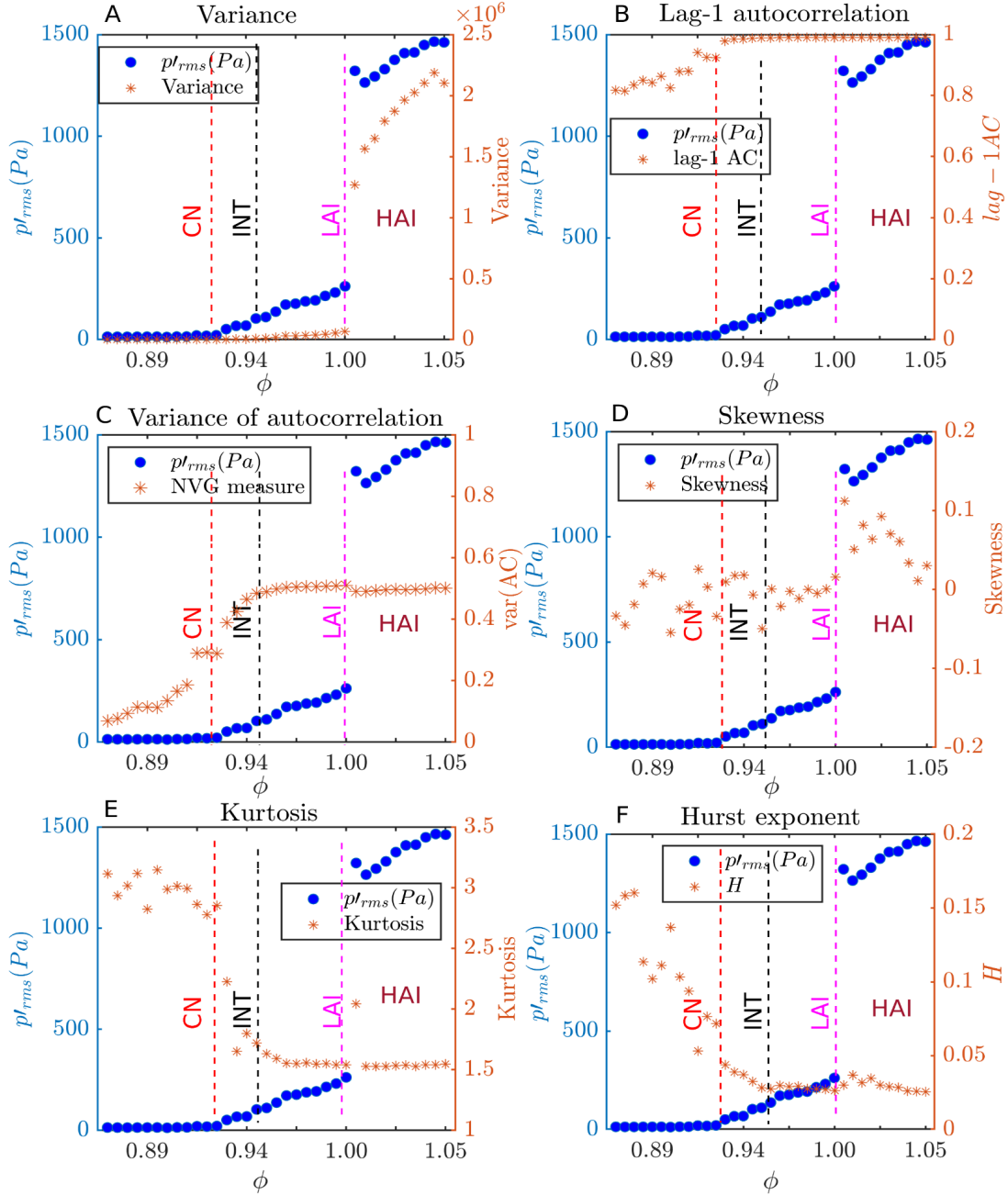

Fig. 1. Performance of traditional early warning measures on the data acquired from the turbulent reacting flow in an annular geometry exhibiting different sequences of bifurcation to oscillatory instability. The blue markers (●) represent the RMS of acoustic pressure fluctuations ($p'_{rms}$) and the orange markers (✶)

represent the different early warning signals (A) Variance, (B) lag-1 autocorrelation, (C) Variance of autocorrelation, (D) Skewness, (E) Kurtosis, and (F) Hurst exponent (*H*). The vertical dashed lines mark the control parameter (equivalence ratio ($\phi$)) at which the system crosses the different states: combustion noise (CN, orange), intermittency (INT, black), and low-amplitude limit cycle oscillations (LAI, magenta). After the LAI line, we have the high-amplitude limit cycle oscillations (HAI). These measures can provide early warning for an impending primary bifurcation, i.e. these measures gradually decrease (E and F) or increase (A–C), and saturate to a value during the primary bifurcation (CN to LAI) and remain the same even for the secondary bifurcation (LAI to HAI).

**Spectral domain-based early warning measures for oscillatory instability**

Previous studies (11,12) have developed early warning signals (EWS) by first converting the time series into the frequency domain, rather than deriving the EWS directly from the time series. Such EWS based on frequency-domain analysis for predicting oscillatory instabilities are shown in Fig. 2. Figure 2A shows the amplitude of the dominant frequency ($A_f$) as the system transitions from safe operation to high-amplitude thermoacoustic instability. The value of the amplitude of the dominant frequency rises in a similar fashion to that of the RMS of acoustic pressure fluctuations and jumps to a higher value at the transition to high-amplitude thermoacoustic instability.

Pavithran *et al.* (17) defined the spectral moment of order two as an early warning signal for forewarning the oscillatory instabilities. Figure 2B shows that the value of the spectral moment of order two gradually decreases (on a log scale) to a value close to zero prior to the primary bifurcation to low-amplitude thermoacoustic instability. Following the onset of low-amplitude thermoacoustic instability, the value of the spectral moment of order two exhibits minimal variation. At the transition to high-amplitude thermoacoustic instability, the value of the spectral moment of order two drops sharply to a lower value of $10^{-2}$ and remains approximately constant throughout the state of high-amplitude thermoacoustic instability. From Fig. 2, we observe that both the amplitude of the dominant frequency and the spectral moment of order two are effective in providing early warning only for an impending primary bifurcation.

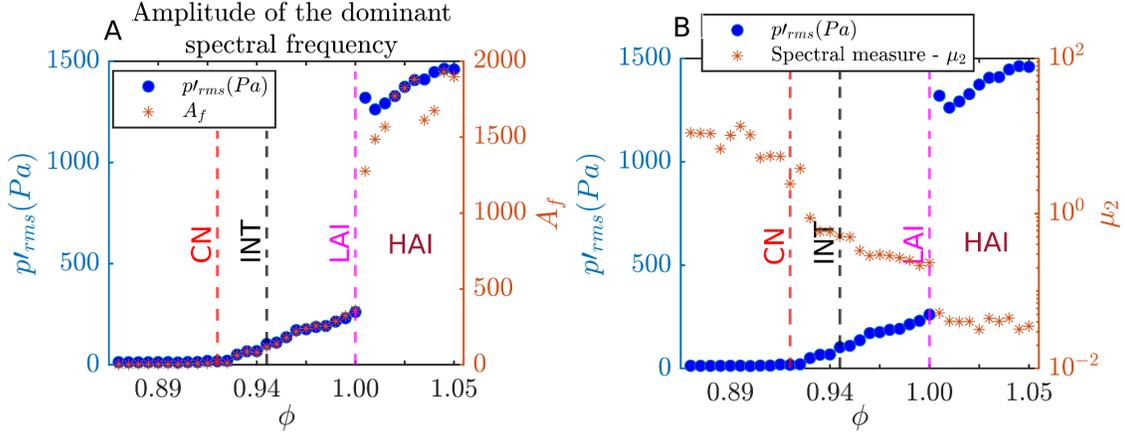

Fig. 2. Working of measures derived from the spectral domain on secondary bifurcation data from the annular combustor. (A) The amplitude of the dominant spectral frequency ($A_f$) varies in a manner similar to the $p'_{rms}$. (B) The second-order spectral moment ($\mu_2$) acts as a precursor to the primary bifurcation, showing a decreasing trend as the system transitions from CN to INT and eventually to LAI. However, the value of $\mu_2$ shows only minimal change prior to the secondary bifurcation (LAI to HAI), before suddenly dropping at the bifurcation from LAI to HAI around $\phi \approx 1$.

In summary, the traditional early warning measures and the spectral-based measures are not sufficient to forewarn an impending primary and secondary bifurcation. Next, we discuss the development of a new early warning measure to forewarn the onset of an impending sequence of oscillatory instabilities in practical systems and a methodology for distinguishing between the primary and the secondary bifurcations. The measure relies on time series data acquired from a variable that reflects the underlying dynamics of the system.

**Visibility graph construction and early warning measure for oscillatory instability**

The choice of the system variable used to develop early warning signals is critical for preventing the system from tipping into undesirable states. As in real engines, the only measurement typically available is data from an acoustic pressure transducer. In the present study, for the annular combustor, we use the time series of acoustic pressure fluctuations for the analysis. Further, with the variable as acoustic pressure fluctuation to derive the early warning measure, we use theory from natural visibility graphs. A natural visibility graph is a tool for encoding the features of a time series into a complex network comprising nodes and links (32 - 34).

Figure 3 illustrates the methodology used for the visibility graph construction from a time series. For example, let us consider the time series of a sine wave with 120 Hz as shown in Fig. 3A. Let the signal $x(N)$ represent $N$ discrete data points of the sine wave. We perform the fast Fourier transform (FFT) of $x(N)$, which converts the signal $x(N)$ to $X(K)$, where $K$ is the frequency index. $X(K)$ is a complex-valued array representing the signal in the frequency domain. We determine the amplitude spectrum $|X(K)|$ corresponding to each frequency as,

$$\text{Amplitude spectrum: } A(K) = |X(K)| = \sqrt{Re(X(K))^2 + Imag(X(K))^2} \qquad [1]$$

Figure 3B shows the amplitude spectrum with Eq. 1 from the acquired time series of acoustic pressure fluctuations. In general, the visibility graph is constructed on the time series of the acquired data. In this study, we construct the visibility graph on the amplitude spectrum (32, 35) as shown in Fig. 3C. The natural visibility graph encodes the frequency modulation characteristics of a uniformly sampled time series (36, 37). In this study, we construct the visibility graph on the frequency domain data, where each frequency value in the amplitude spectrum is considered to be a node. Two nodes are connected if a straight line drawn between their corresponding data points does not intersect any intermediate point, meaning the nodes are mutually 'visible'. From a mathematical perspective, nodes $i$ and $j$ of time series $X(K)$ share a connection when every intervening node $n$ meets the condition:

$$X(K_n) < X(K_j) + (X(K_i) - X(K_j))(K_j - K_n)/(K_j - K_i) \qquad [2]$$

In our study, we consider the node with the maximum amplitude in the amplitude spectrum (representing the dominant frequency of the data) as our reference node. While constructing the visibility graph (35), connections with other nodes are made only if they are directly visible from the reference node (see Fig. 3C).

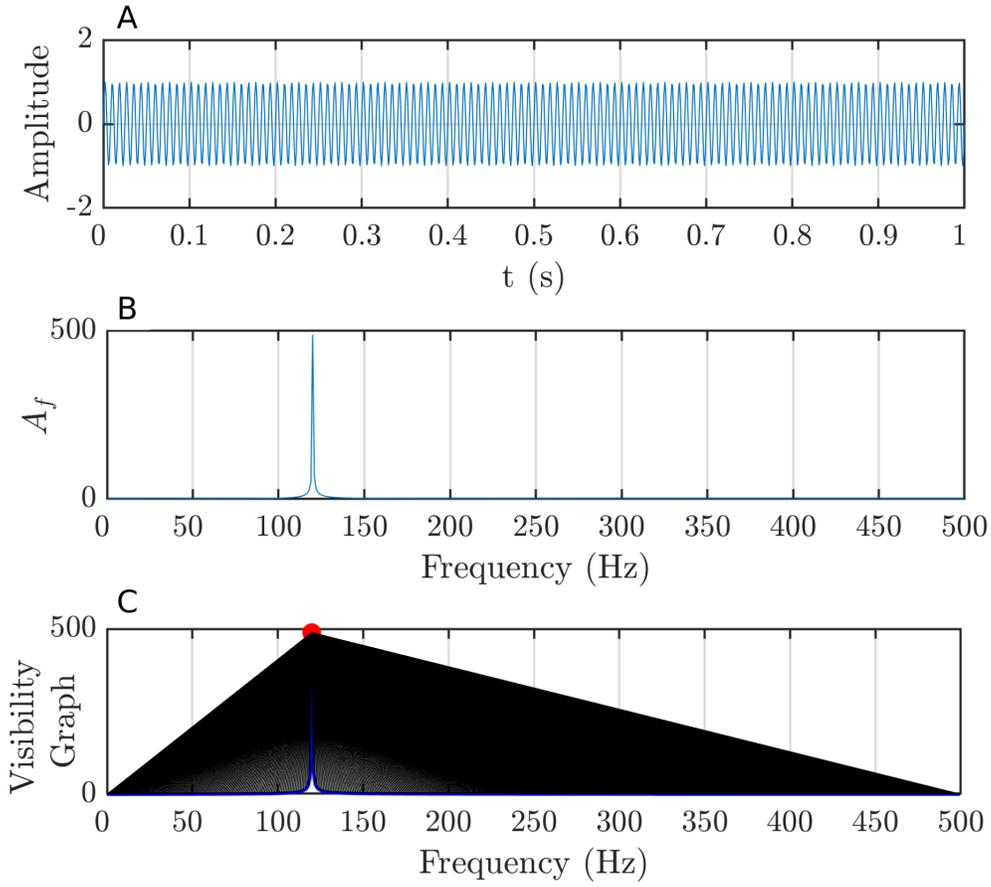

Fig. 3. Schematic of the construction of the visibility graph for a given time series signal. (A) A sample time series of a 120 Hz sine wave. (B) The amplitude ($A_f$) spectrum is plotted against the frequencies in the signal (shown only up to 500 Hz data here), showing a prominent peak at 120 Hz. (C) The visibility graph is constructed using the peak node ● in the amplitude spectrum as the reference, with connections displayed to all other frequency components.

For extracting early warning signals, we define a term called natural visibility graph measure (NVGM) as,

$$\text{NVGM} = 1 - \frac{\text{Degree of the reference node in the amplitude spectrum}}{\text{Maximum degree possible for the reference node}} \quad [3]$$

The NVGM quantifies the lack of visibility of all other nodes (i.e., frequency components) from the reference node in the signal being analyzed. An NVGM value of zero indicates that the reference node can see all other nodes in the amplitude spectrum, suggesting that a single frequency is dominating in the signal. Conversely, when the value of NVGM is close to 1, it suggests very low visibility for the reference node, indicating a broad range of frequency components in the signal and suggesting an aperiodic or chaotic state. We use this NVGM as an early warning signal to predict an impending oscillatory instability, where the NVGM shifts from a value close to 1 to a value close to 0 as the system transitions from a disordered to an ordered state.

Before applying this method, it is essential to understand what the NVGM representation encodes and why it provides richer information than conventional signal-analysis tools. Time-domain observations allow one to view waveform amplitude and temporal evolution directly, but many of the most informative dynamical features, such as oscillatory components, harmonics, and the distribution of energy across scales, are not easily distinguished from broadband fluctuations (35). This limitation becomes more pronounced in systems characterized by multi-scale dynamics, where time-domain indicators often fail to isolate the underlying structure of the signal.

Frequency-domain representations show this intrinsic structure more effectively by decomposing the signal into spectral components that directly reflect its dynamical content. Visibility-graph methods provide a complementary viewpoint (35) by converting a temporally ordered sequence into a network whose edges encode geometric visibility relations. They capture nonlinearities, scale interactions, and structural patterns that are difficult to infer from the original signal alone. Visibility graphs preserve key properties such as periodicity and exhibit invariance under several transformations, since they depend only on the relative ordering and spacing of data values rather than their absolute magnitudes (33). These features have made visibility-graph approaches powerful tools for analyzing complex time series.

Here, we extend this framework to amplitude and power spectra, constructing visibility graphs on the spectral magnitude and power sequence rather than the raw temporal signal. This preserves the desirable invariances of the visibility mapping while enabling the graph structure to highlight the underlying structure in the frequency domain that may not be apparent in the amplitude spectrum itself. In a typical spectrum, narrowband peaks correspond to harmonic or resonant activity, whereas broadband energy arises from turbulent fluctuations. In thermoacoustic systems, such narrowband contributions often originate from the coupled response of flow, flame, and acoustic fields, potentially overwhelming weaker harmonic features.

A key advantage of the spectral visibility-graph degree is its ability to emphasize salient harmonic peaks even when substantial broadband energy is present. The visibility graph representation, as shown in Fig. 3C, incorporates information about the local spectral environment of each point. Peaks that are well separated from neighbouring features map to nodes with a higher degree, whereas closely spaced peaks produce comparatively lower values of degree. The dependency on the position of the peak acts as a form of nonlinear compression, enhancing sparse harmonic content while suppressing clutter arising from broadband components. As a result, the NVGM offers a robust route for revealing the hidden structure in complex spectral data and for isolating dynamical signatures that conventional spectral representations may not capture.

In practical scenarios, systems may follow different sequences of bifurcation. These may include a primary bifurcation that is either continuous or abrupt (1), a primary followed by an abrupt secondary bifurcation (24, 29, 31). From Eq. 1, we get the amplitude spectrum, over which we constructed the visibility graph. Further, we can also use the power spectrum to construct a visibility graph. The power spectrum represents the power contribution of each discrete frequency bin, defined as,

Power spectrum: $P(K) = |X(K)|^2 = Re(X(K))^2 + Imag(X(K))^2$     [4]

We further generalize this by defining the spectrum on which the visibility graph will be constructed as,

Generalized spectrum of order $q$: $G(K) = |X(K)|^q$     [5]

where $q$ represents the sensitivity parameter. Here, $q$ can be considered analogous to the $q$ factor in multifractality, where we give relative importance to different amplitude fluctuations. Here, we focus on different amplitude peaks in the spectrum. By increasing/decreasing $q$, we are enhancing those small peaks and looking for a very clean periodic signal.

To forewarn an impending sequence of bifurcations, as a general procedure, we start with the time series data obtained from the system. In the case of the annular combustor, we have the time series of acoustic pressure fluctuations. Further, we convert the data from the time domain into the frequency domain using the Fast Fourier Transform (FFT). Spectral analysis was performed on 1s long Hann-windowed pressure segments, yielding a true frequency resolution of 1 Hz. The signal was zero-padded prior to FFT computation, resulting in a refined spectral grid with a frequency bin spacing of approximately 0.61 Hz. An uncertainty analysis with the change in window size is presented in the *Methods* section.

As shown in Fig. 4A(i) – D(i), when the sensitivity parameter is set to $q = 2$, the power spectrum is calculated according to Eq. 4. In this case, the value of spectral power at the peak node is significantly higher than that of the other frequency components, leading to enhanced visibility of the dominant frequency relative to all other components. As the system approaches the state of intermittency, a single dominant peak emerges, and the visibility of the peak node reaches its maximum near the state shown in Fig. 4B(i). Beyond this point, in Fig. 4C(i) – 4D(i) in the state of low and high-amplitude thermoacoustic instability, a single frequency dominates.

In Fig. 4A(ii) – 4D(ii), for the same time series of acoustic pressure fluctuations, we set the sensitivity parameter to $q = 1$. We observe that the value of the amplitude of the dominant peak relative to other frequency components is lower compared to the case of $q = 2$, leading to reduced visibility of the peak frequency across Fig. 4A(ii) – B(ii). As the system transitions to the state of intermittency, a single dominant peak still emerges. However, the visibility of the peak node reaches its maximum only when the system exhibits a sharp peak in the amplitude spectrum, near the state of low-amplitude thermoacoustic instability, as shown in Fig. 4C(ii). Thereafter, in the state of low-amplitude thermoacoustic instability and high-amplitude thermoacoustic instability as shown in Fig. 4C(ii) – 4D(ii), a single frequency becomes increasingly dominant.

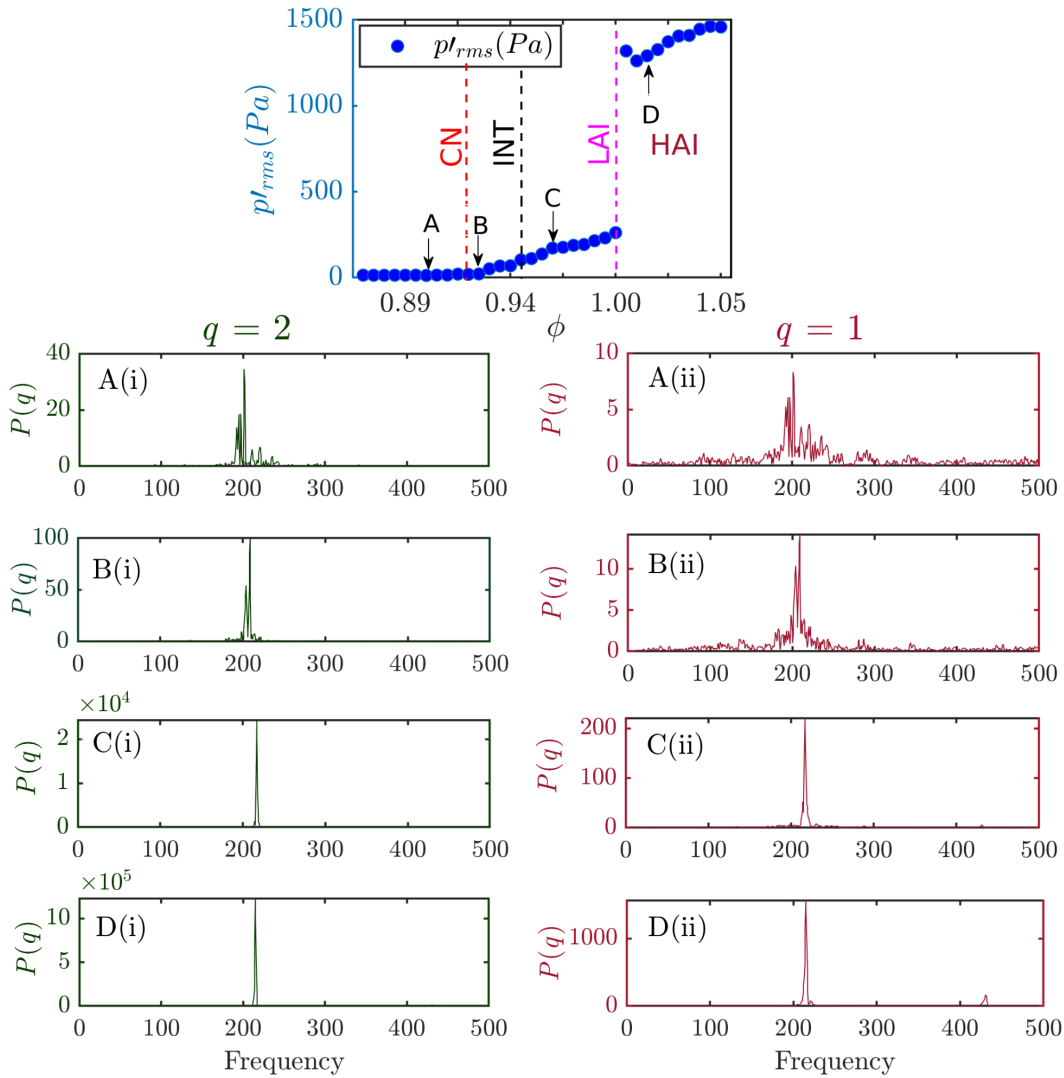

Fig. 4. Amplitude ($q = 1$) and power ($q = 2$) spectrum plots corresponding to different states observed during different sequences of bifurcations to oscillatory instabilities. The top figure displays the bifurcation plot derived from the time series of acoustic pressure fluctuations from an annular combustor, with the different observed states labeled as A (CN), B (INT), C (LAI), and D (HAI). A(i) - D(i), in green colour illustrates the power spectrum ($q = 2$) corresponding to states A, B, C, and D. A(ii) - D(ii), in red shows the amplitude spectrum ($q = 1$) corresponding to states A, B, C, and D.

We quantify the relative difference in visibility, i.e., we emphasise different amplitude fluctuations in the spectrum by varying $q$. When the small peaks are amplified or reduced, it affects the visibility of the dominant peak. This allows us to provide early warnings of both primary and secondary bifurcations. The parameter $q$, referred to as the sensitivity parameter, effectively captures the sensitivity of the signal, i.e., it quantifies the amplification/diminution of some of the peaks in the spectrum. We compute the NVGM$|_q$ for different values of $q$, as shown in Fig. 5.

When $q = 2$, the value of NVGM$|_{q=2}$ has high sensitivity, meaning that as the system approaches the oscillatory state, the dominant peak rises rapidly compared to all other frequency components in the signal. This growth is effectively captured by NVGM$|_{q=2}$. A value of 0.1 is chosen as the threshold for the current system to provide an early warning of impending bifurcations. The choice of threshold depends on the characteristics of the system and the risk-taking tolerance at the operator end. As a general guideline, the threshold should be set to minimize false alarms while maximizing the available warning time. As an example, we consider that an early warning is triggered when the value of NVGM$|_{q=1,2}$ drops below the value of 0.1. For $q = 1$, the behavior of NVGM$|_{q=1}$ is shown in Fig. 5B. In this case, NVGM$|_{q=1}$ remains relatively high during the state of safe operation and intermittency states and drops below 0.1 only during the state of low-amplitude thermoacoustic instability. Thus, it provides an early warning of an impending secondary bifurcation to oscillatory instability, marking the transition from low-amplitude to high-amplitude thermoacoustic instability.

While previous measures reported in the literature tend to saturate after the primary bifurcation (Fig. 1), NVGM$|_q$ is capable of providing early warnings for both primary and secondary bifurcations simply by adjusting the value of $q$ from 2 to 1. We refer to this methodology, where the time series of acoustic pressure fluctuations is analyzed twice: once to forewarn of the primary bifurcation ($q = 2$) and once for the secondary bifurcation ($q = 1$), as the concept of staging. Essentially, assigning a higher value to $q$ increases the sensitivity of our EWS to detect any emerging peak at the earliest opportunity. This staged approach can also be extended to provide early warnings for any subsequent transitions beyond the secondary bifurcation.

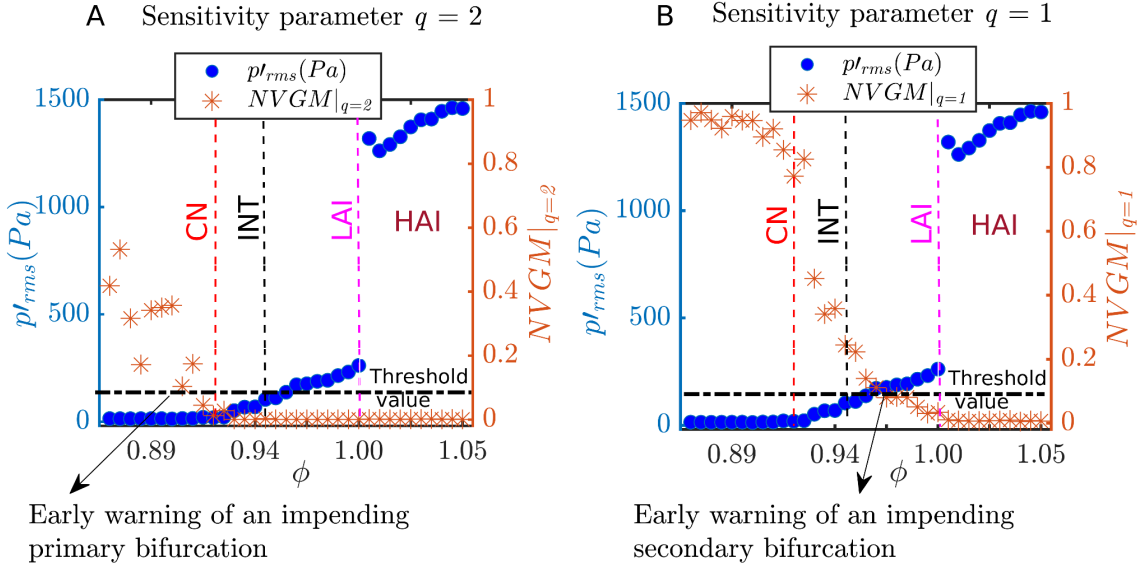

Fig. 5. Staging concept for obtaining early warnings of an impending primary and secondary bifurcation to oscillatory instabilities. The blue marker (●) represents the RMS of acoustic pressure fluctuations ($p'_{rms}$) in the bifurcation plot derived from the time series of acoustic pressure fluctuations of an annular combustor. The orange marker (✶) denotes the value of $\text{NVGM}|_{q=1,2}$ with a value of 0.1 as the threshold represented by the dotted black line. (A) For $q = 2$, the value of $\text{NVGM}|_{q=2}$ is calculated from the power spectrum of the signal. The value of $\text{NVGM}|_{q=2}$ drops below the value 0.1, as indicated by the arrow, providing an early warning of an impending primary bifurcation, i.e., transition to LAI. (B) For $q = 1$, the value of $\text{NVGM}|_{q=1}$ is calculated on the amplitude spectrum derived from the signal. The $\text{NVGM}|_{q=1}$ decreases gradually, well before the rise in $p'_{rms}$ to high amplitudes. The value of $\text{NVGM}|_{q=1}$ decreases to a value below 0.1, shown by the arrow, thus providing an early warning of an impending secondary bifurcation (transition from LAI to HAI).

We demonstrate the effectiveness of our methodology and the early warning signal $\text{NVGM}|_q$ by analyzing time series data from various systems, including secondary bifurcation data from a backwards-facing step combustor with a swirler and bluff body configuration, as well as from an aeroacoustic system, as shown in Fig. 6(A - F). A detailed description of the experimental setup and the operating conditions is mentioned in the *Methods* section.

Figure 6(A, B) illustrates the variation of the control parameter (equivalence ratio) against the RMS of acoustic pressure fluctuations acquired from a dump combustor with a bluff-body as the flame-holding mechanism. A value of 0.12 is considered as the threshold for $\text{NVGM}|_{q=1,2}$ effectively provides early warnings of impending primary and secondary bifurcations to oscillatory instability in the bluff-body-stabilized dump combustor. This trend is consistent with traditional early-warning measures (11), which also require lower threshold values to reliably detect warnings in turbulent combustors. The FFT used for NVGM estimation is evaluated with a Hann window applied over the full 1 s data segment, yielding a true frequency resolution of 1 Hz. Zero-padding is employed to refine the spectral bin spacing. A frequency bin spacing of ~0.6 Hz is chosen for this system. For $q = 2$, the value of $\text{NVGM}|_{q=2}$ drops below the threshold near

$\phi = 1.97$. When using $q = 1$, the values of NVGM$|_{q=1}$ remain elevated in the initial stages, where NVGM$|_{q=2}$ is near zero, indicating the potential for a further transition. Consistent with this prediction, the $p'_{rms}$ rises sharply at $\phi = 1.59$.

Similarly, Fig. 6(C, D) shows the bifurcation diagram along with the variation of NVGM$|_{q=1,2}$ in a dump combustor with a swirler serving as the flame-holding mechanism. A threshold value of 0.05 for NVGM$|_{q=1,2}$ provides effective early warning of impending primary and secondary bifurcations to oscillatory instability in the swirl-stabilized combustor. In turbulent combustors, the observed dynamics emerge from strong interactions among multiple subsystems, which include the turbulent flow field, unsteady heat release, and acoustic field. Consequently, the pressure measurements inherently reflect the combined influence of these coupled processes. Due to this complex, multiscale coupling, the magnitude of fluctuations associated with early-warning signals is comparatively small, necessitating the use of lower threshold values for reliable detection. The FFT used for NVGM estimation is evaluated with a Hann window applied over the full 1 s data segment, yielding a true frequency resolution of 1 Hz. Zero-padding is employed to refine the spectral bin spacing. A frequency bin spacing of ~0.6 Hz is chosen for this system. For $q = 2$, the value of NVGM$|_{q=2}$ drops below the threshold near $\phi = 1.3$. The values of NVGM$|_{q=1}$ remain high in the initial stages, where NVGM$|_{q=2}$ is near zero, indicating the potential for a further transition. Consistent with this prediction, the $p'_{rms}$ rises sharply at $\phi = 1.18$.

We further apply our methodology to the onset of aeroacoustic oscillations. An aeroacoustic system with confined flow through the double orifices is chosen for the present study. Figure 6(E, F) shows the variation of $p'_{rms}$ as a function of the Reynolds number ($Re$), ranging from 2600 - 2200, and we observe different jumps in the $p'_{rms}$. The FFT used for NVGM estimation is evaluated with a Hann window applied over the full 0.25 s data segment, yielding a true frequency resolution of 4 Hz. Zero-padding is employed to refine the spectral bin spacing. A frequency bin spacing of ~0.8 Hz is chosen for this system. For $q = 2$, the value of NVGM$|_{q=2}$ gradually decreases and drops below the value of threshold = 0.3, nearly at $Re \approx 2480$. When we use $q = 1$, the values of NVGM$|_{q=1}$ remain high in the initial stages, where NVGM$|_{q=2}$ is near zero, indicating the potential for a further transition. Consistent with this prediction, $p'_{rms}$ rises sharply at $Re \approx 2200$. To obtain an early warning of the impending secondary jump, a value of 0.3 is chosen as the threshold, yielding a warning at $Re \approx 2240$.

Thus, system-specific window selection ensures comparable fidelity in capturing the relevant spectral features governing the estimation of NVGM across configurations with inherently different dynamical regimes. The choice of threshold ultimately depends on the acceptable level of risk and the characteristics of the system. As a general guideline, it is advisable to begin with a value of nearly 0.2 - 0.3 as the threshold and refine it based on system-specific behaviour. Overall, the natural visibility graph-based measure has proven effective in offering early warnings for both primary and secondary bifurcations across different practical systems.

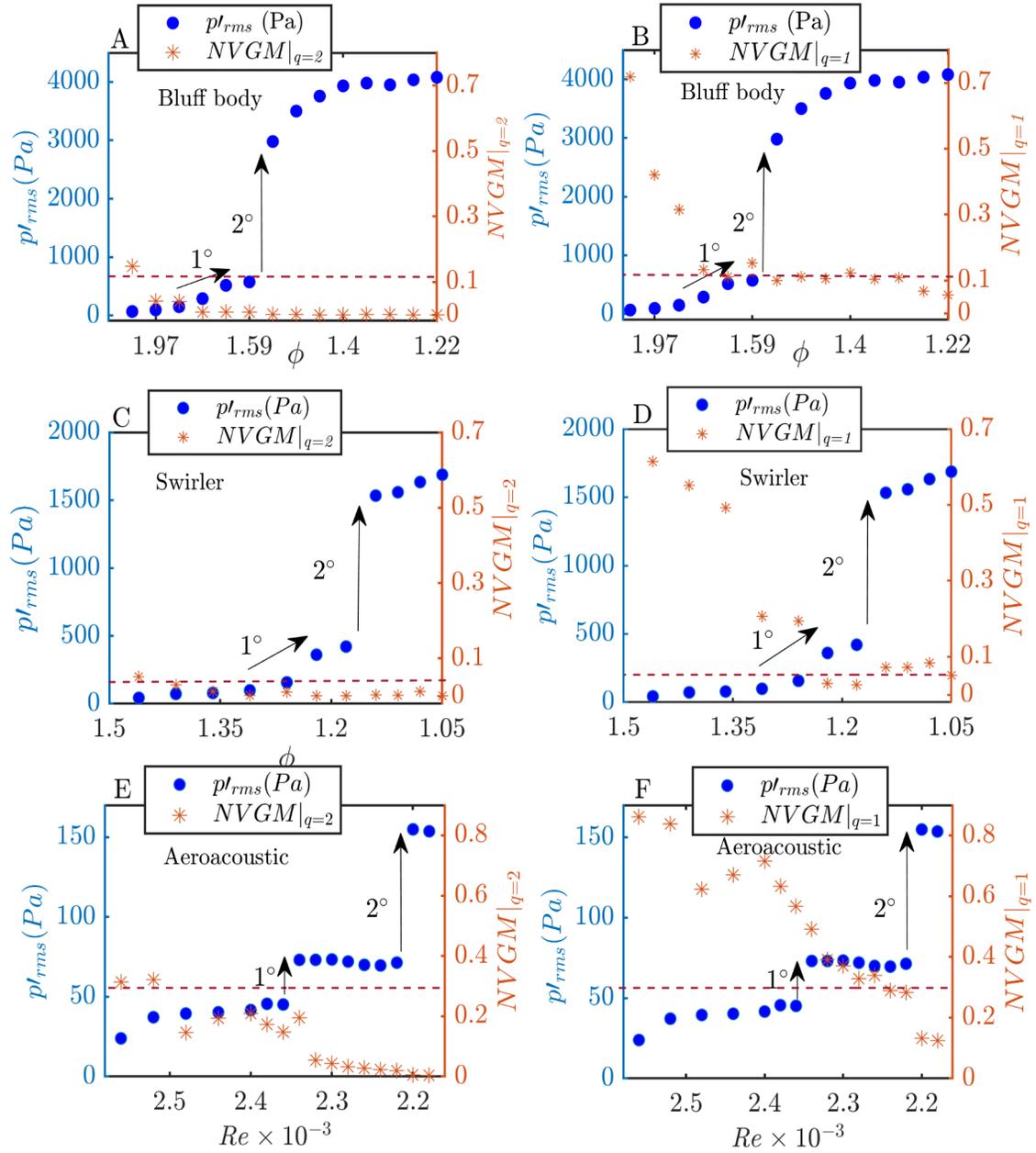

Fig. 6. Secondary bifurcation data from different practical systems demonstrating the effectiveness of the early-warning indicator NVGM|$_q$ in forewarning primary and secondary transitions to oscillatory instability. (A, B) Bifurcation diagram showing the variation of the RMS of acoustic pressure fluctuations with the control parameter (equivalence ratio), along with the corresponding evolution of NVGM|$_{q = 1, 2}$ for a dump combustor with a bluff-body configuration. (C, D) Similar results for a dump combustor with a swirler used for flame stabilization. (E, F) Bifurcation diagram for an aeroacoustic system showing the variation of the RMS of acoustic pressure fluctuations with Reynolds number, together with NVGM|$_{q = 1, 2}$. Values of 0.12 (bluff body), 0.05 (swirler), and 0.3 (aeroacoustic system) are used as thresholds to issue early warnings in their respective systems.

**Discussion**

We validated our methodology for forewarning different sequences of bifurcations leading to oscillatory instabilities. The current early warning signal, based on the natural visibility graph applied to the frequency domain, is effective in detecting transitions to oscillatory instabilities. This is achieved since the frequency domain effectively captures the emergence of periodic behavior in the system.

While several early warning signals are available for predicting oscillatory instabilities, in some cases, forewarning the primary bifurcation alone is sufficient, since certain systems go to undesirable states at these low amplitudes associated with a primary bifurcation. The methodology presented here offers early warning of such primary bifurcations, making it comparable in function to existing early warning signal approaches. For systems that can tolerate low-amplitude oscillations, the concept of staging can be particularly useful. By initially using $q = 2$ to detect the onset of a primary bifurcation, and subsequently analyzing the same data with $q = 1$, we can predict the possibility of further transitions, without disrupting the system operation. Such prediction capability is beneficial for optimizing the system performance while avoiding tipping into undesirable states.

The effectiveness of receiving early warning signals depends on the availability of high-resolution data leading up to the tipping point (38). Furthermore, while we can forewarn an impending primary or secondary bifurcation, we cannot precisely predict the exact moment of transition. It is important to note that receiving an early warning signal alone does not guarantee the prevention of a system tipping into an undesirable state. Effective prevention also requires appropriate actions based on system conditions and variations in control parameters, alongside accurate early warning measures. The selection of the value of the threshold in our methodology varies depending on the system under study. Setting an appropriate threshold requires prior knowledge of the characteristics of the system to effectively apply the approach presented in this work.

While other early warning signals are available that primarily forewarn only the onset of primary bifurcations leading to oscillatory instability, our approach offers a new pathway to identify precursors across different sequences of bifurcations. Moreover, it demonstrates a high degree of generality across various systems exhibiting transitions to oscillatory states. This insight is valuable for both theoretical understanding and practical applications, as tipping points in many systems can result in undesirable collapses (39). Enhanced early warning capabilities can significantly improve our ability to prevent or better prepare for such critical transitions (40).

**Acknowledgments**

We thank Dr. Samarjeet Singh, Dr. Amitesh Roy, Dr. Manikandan Raghunanthan, and Dr. Ramesh Bhavi for their experimental data. RR acknowledges the support from the Prime Minister Research Fellowship, Government of India. RIS acknowledges the funding from the IoE initiative (SP22231222CPETWOCTSHOC) and the Department of Science and Technology, Government of India (SERB/CRG/2020/003051).

**Author Contributions:** Conceptualization: RR, PK, IP, RIS; Data curation: RR; Formal analysis: RR, PK; Investigation: RR, PK, RIS; Methodology: RR, PK, IP, RIS; Supervision: IP, RIS; Validation: RR; Visualization: RR, PK, IP; Funding acquisition: RIS; Project administration: RIS; Writing – original draft: RR; Writing – review & editing: PK, IP, RIS

**Competing Interest Statement:** The Authors declare that they have no competing interests.

## Methods

The following section outlines the experimental setup and operating conditions of the various turbulent combustors and the aeroacoustic system used in this study.

### Instrumentation

In all the thermofluid systems used in this study, which include annular combustor, bluff-body, and swirl stabilized turbulent combustors, the flow rates of air and fuel are controlled using mass flow controllers (Alicat mass flow controllers, MCR series) with a measurement uncertainty of ±(0.8% of reading + 0.2% of full scale). In all experiments, the control parameter, which is the equivalence ratio ($\phi$), varied in a quasi-static manner. The acoustic pressure fluctuations are acquired using a piezoelectric pressure transducer (make: Piezotronics, model: PCB103B02) at a sampling rate of 10 kHz for a duration of 3 s. The maximum uncertainty in the reported values of pressure measurement is ±0.15 Pa.

**Bluff-body stabilized turbulent combustor:** Figure 7(a, c) presents the schematic of the combustor, which consists of a plenum chamber, a combustion chamber, and an acoustic decoupler. The central shaft supporting the bluff body also serves as the fuel delivery duct, supplying fuel through four radially arranged injection holes. The injection takes place just before the dump plane. A bluff body with a diameter of 47 mm and 10 mm thickness is used to stabilize the flame. The combustor has a square cross-section with dimensions 90 × 90 mm² and an overall length of 1400 mm. In the experiments, the equivalence ratio ($\phi$) is systematically varied while maintaining a constant fuel flow rate of 48 SLPM. The airflow rate is increased from 700 SLPM to 1232 SLPM, and the corresponding equivalence ratio changes from $\phi$ = 2.10 to 1.22. The Reynolds number calculated based on the bluff-body diameter ranges from 3.84 × 10$^4$ to 6.1 × 10$^4$. For more details on the experimental setup, please refer to (16, 30).

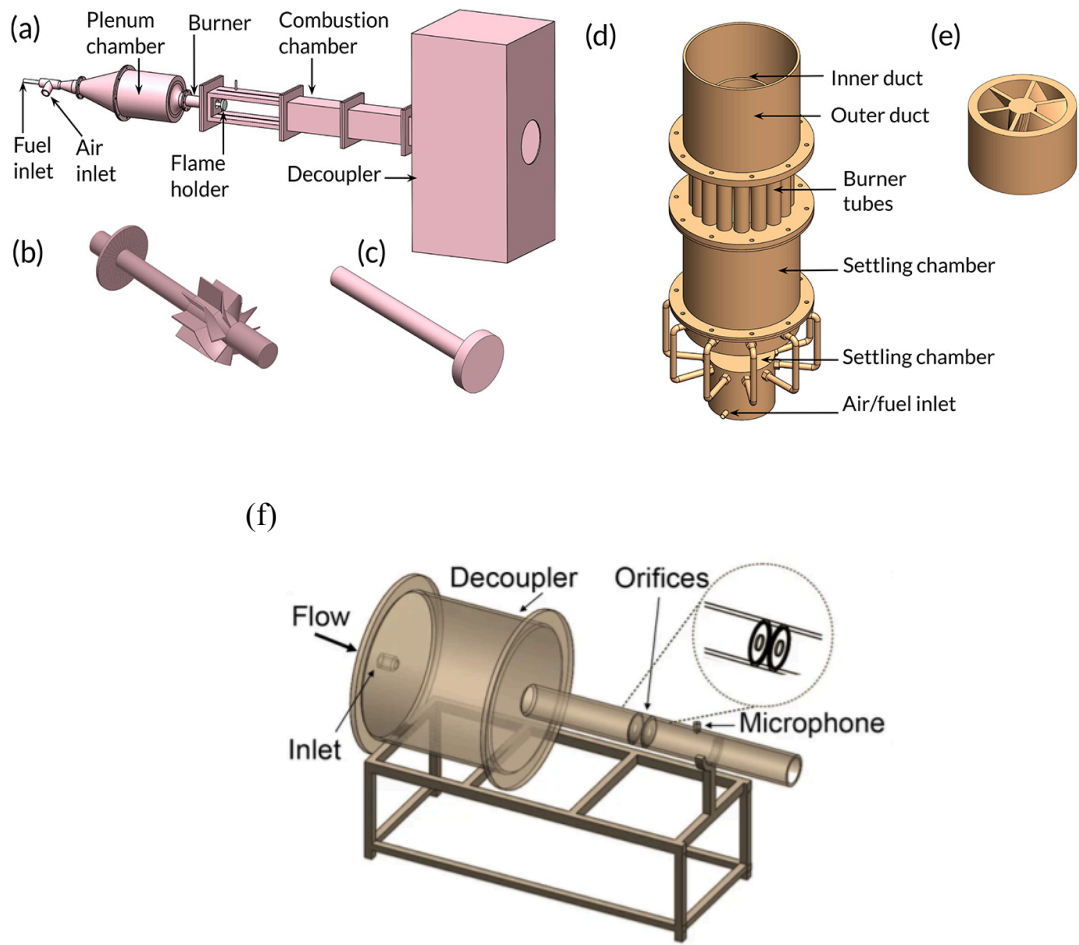

Fig. 7. Schematic of the practical systems studied in forewarning primary and secondary bifurcation to oscillatory instability includes: (a, b) a dump combustor with a swirler configuration, (a, c) a dump combustor with a bluff body configuration. (d, e) shows the schematic of an annular combustor with sixteen swirl-stabilized burners. These figures are reproduced with permission from (41) Copyright© 2024 American Institute of Physics. (f) Aeroacoustic system, reproduced from (17) Copyright© 2020, licensed under CC BY 4.0.

**Swirl-stabilized turbulent combustor:** The schematic of the swirl-stabilized combustor is shown in Fig. 7(a, b). The method of injecting the fuel-air mixture is similar to that used in the bluff-body stabilized configuration. In this setup, a technically premixed air - LPG mixture flows through a swirler (diameter: 40 mm) containing eight vanes, each inclined at 40° to the longitudinal axis, as shown in Fig. 7(b). The swirler is used for flame stabilization and is positioned such that the trailing edges of the vanes are flush with the dump plane. The equivalence ratio ($\phi$) is changed systematically by increasing the airflow rate from 380 SLPM to 700 SLPM in steps of 20 SLPM, while keeping the fuel flow rate fixed at 24 SLPM. The airflow rate is varied in a quasi-static manner, such that the value of $\phi$ decreases from 1.5 to 1.05. The corresponding thermal power of the combustor, based on the constant fuel flow rate, is 24.30 kW. The Reynolds number, calculated using the swirler diameter, spans from $Re_d = 1.7 \times 10^4$ and $2.9 \times 10^4$. For additional details regarding the experimental setup, please refer to (30).

**Annular combustor:** Figure 7(d, e) shows the schematic of the swirl-stabilized annular combustor used in the experiments. A technically premixed mixture of air and liquefied petroleum gas (LPG, composed of 40% propane and 60% butane) is used as the fuel for the experiments. The air-fuel mixture enters through an inlet connected to a settling chamber with a flow straightener inside the chamber to minimize flow irregularities and a hemispherical flow divider. The mixture from the settling chamber feeds into 16 burners arranged in an annular configuration. Each burner tube has a diameter of 30 mm and a length of 150 mm. These burners lead into the combustion chamber, which consists of concentric inner and outer ducts measuring 200 mm and 400 mm in length, respectively. Each burner uses an internal swirler to stabilize the flame. Each swirler contains six vanes set at an angle of $\beta = 60°$ relative to the burner axis, as shown in Fig. 7(e).

Ignition is initiated with a non-premixed pilot flame, which is turned off once the flame is stabilized. Throughout the experiments, the air flow rate is maintained at 1400 SLPM, and the fuel flow rate is increased from 40 to 48 SLPM, such that the equivalence ratio ($\phi$) increases from 0.8 to 1.1. Using the burner exit diameter of 15 mm as the characteristic length, the corresponding Reynolds number is approximately 8600. With reference to the fuel flow rate, the thermal power output of the combustor ranges from 39 to 79 kW. For further details regarding the time series of acoustic pressure fluctuations at various equivalence ratios, please refer to (29 - 31).

**Aeroacoustic system:** We present the schematic of the experimental setup in Fig. 7(f). Air enters the plenum chamber through an inlet port, which is connected to a circular duct measuring 613 mm in length and 50 mm in diameter. This duct accommodates two orifices, each with a diameter of 20 mm and a thickness of 2.5 mm. The orifices are spaced 18 mm apart, with the first orifice positioned 300 mm downstream from the plenum chamber. The experiments are performed at a room temperature of 27°C. Reynolds number ($Re$) is the control parameter in this system, which is varied systematically and the corresponding velocity fluctuations and acoustic pressure oscillations are acquired. $Re$ is calculated using the expression $Re = (\bar{\varrho} u l_c)/\mu$, where $u$ is the bulk horizontal velocity at the orifice, $\bar{\varrho}$ is the air density (kg/m³), $l_c$ is the characteristic length (taken as the orifice diameter), and $\mu$ is the dynamic viscosity of air.

An Alicat MCR-series mass flow controller (MFC) is used to vary the air flow rate, which has a measurement uncertainty of $\pm(0.8\%$ of the reading $+ 0.2\%$ of the full-scale value). The airflow rate is varied in a quasi-static manner from 224 SLPM to 178 SLPM, corresponding to Reynolds numbers ranging from nearly $2600 \pm 10$ to $2200 \pm 10$.

A pre-polarized pressure field microphone (Piezotronics PCB378C10) is mounted on the pipe wall, 305 mm downstream from the plenum chamber, to measure acoustic pressure oscillations. This microphone offers a sensitivity of 1 mV/Pa and a resolution of 20 µPa. Acoustic pressure is acquired over a duration of 5 s at a sampling rate of 20 kHz. Please refer to (30) for any further information regarding the setup.

**Assessing the influence of spectral resolution on NVGM uncertainty**

Figure 8 presents the values of $\text{NVGM}|_{q = 1, 2}$ computed from the spectral data of the annular combustor. The amplitude and power spectrum are evaluated with a window duration of 1s with 10,000 samples, corresponding to a true frequency resolution of 1 Hz. For a fixed frequency resolution, the number of FFT points is varied through zero-padding to modify the frequency bin spacing. A comparison across frequency bin spacing of 1, 0.61, 0.31, and 0.15 Hz shows that the calculated $\text{NVGM}|_{q = 1, 2}$ has weak sensitivity to chosen frequency bin spacing. The values of $\text{NVGM}|_{q = 1, 2}$ show minor variations until 0.31 Hz frequency bin spacing. Accordingly, a frequency bin spacing of 0.61 Hz is adopted in the present study.

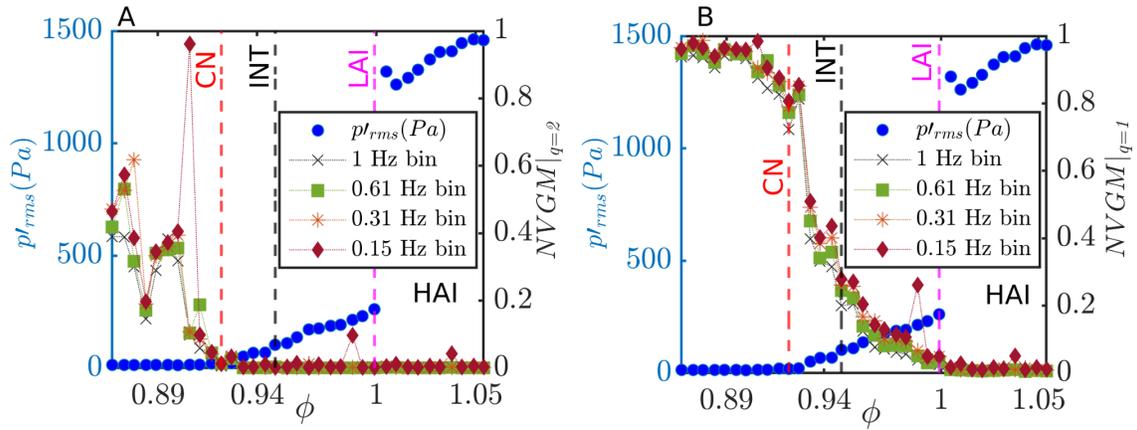

Fig. 8. (A) $\text{NVGM}|_{q = 2}$ and (B) $\text{NVGM}|_{q = 1}$ computed for frequency bin spacing of 1, 0.61, 0.31, and 0.15 Hz from the power and amplitude spectrum of the annular combustor data. The results show that NVGM exhibits minimal sensitivity to the choice of frequency bin spacing.